\documentclass[preprint, superscriptaddress, amsfonts, amsmath]{revtex4-2}

\usepackage[ruled]{algorithm2e}
\usepackage{graphicx}
\usepackage{multirow}

\usepackage{xcolor}

\begin{document}
\title{A machine learning algorithm for direct detection of axion-like particle domain walls}

\author{Dongok Kim}
\affiliation{Department of Physics, Korea Advanced Institute of Science and Technology, Republic of Korea, 34141}
\affiliation{Center for Axion and Precision Physics Research, Institute for Basic Science, Republic of Korea, 34051}

\author{Derek F. Jackson Kimball}
\affiliation{Department of Physics, California State University — East Bay, Hayward, CA 94542-3084, USA}

\author{Hector Masia-Roig}
\affiliation{Johannes Gutenberg-Universit\"at Mainz, 55128 Mainz, Germany}
\affiliation{Helmholtz Institut Mainz, Johannes Gutenberg-Universit\"at, 55099 Mainz, Germany}

\author{Joseph A. Smiga}
\affiliation{Johannes Gutenberg-Universit\"at Mainz, 55128 Mainz, Germany}
\affiliation{Helmholtz Institut Mainz, Johannes Gutenberg-Universit\"at, 55099 Mainz, Germany}

\author{Arne Wickenbrock} 
\affiliation{Helmholtz Institut Mainz, Johannes Gutenberg-Universit\"at, 55099 Mainz, Germany}
\affiliation{Johannes Gutenberg-Universit\"at Mainz, 55128 Mainz, Germany}

\author{Dmitry Budker}
\affiliation{Helmholtz Institut Mainz, Johannes Gutenberg-Universit\"at, 55099 Mainz, Germany}
\affiliation{Johannes Gutenberg-Universit\"at Mainz, 55128 Mainz, Germany}
\affiliation{Department of Physics, University of California, Berkeley, CA 94720-7300, USA}

\author{Younggeun Kim}
\affiliation{Department of Physics, Korea Advanced Institute of Science and Technology, Republic of Korea, 34141}
\affiliation{Center for Axion and Precision Physics Research, Institute for Basic Science, Republic of Korea, 34051}

\author{Yun Chang Shin}
\email{yunshin@ibs.re.kr}
\affiliation{Center for Axion and Precision Physics Research, Institute for Basic Science, Republic of Korea, 34051}

\author{Yannis K. Semertzidis}
\affiliation{Center for Axion and Precision Physics Research, Institute for Basic Science, Republic of Korea, 34051}
\affiliation{Department of Physics, Korea Advanced Institute of Science and Technology, Republic of Korea, 34141}

\date{\today}

\begin{abstract}
The Global Network of Optical Magnetometers for Exotic physics searches~(GNOME) conducts an experimental search for certain forms of dark matter based on their spatiotemporal signatures imprinted on a global array of synchronized atomic magnetometers. The experiment described here looks for a gradient coupling of axion-like particles (ALPs) with proton spins as a signature of locally dense dark matter objects such as domain walls. In this work, stochastic optimization with machine learning is proposed for use in a search for ALP domain walls based on GNOME data. The validity and reliability of this method were verified using binary classification. The projected sensitivity of this new analysis method for ALP domain-wall crossing events is presented.
\end{abstract}
\maketitle

\newpage

\section{Introduction}
Even though there is a considerable amount of evidence for the existence of dark matter, the nature of dark matter is not fully understood~\cite{Read_2014}. Up to now, there have been a number of hypotheses proposed to explain the existence of dark matter~\cite{JHEGYI198328,PRESKILL1983127,ABBOTT1983133,DINE1983137,PhysRevD.30.272,PhysRevLett.72.17,JUNGMAN1996195,PhysRevLett.85.1158,RevModPhys.82.557}.
One of the most well-motivated dark matter candidates is an axion, which was originally proposed to solve the strong CP problem in quantum chromodynamics~(QCD)~\cite{1977PhRvL..38.1440P,1977PhRvD..16.1791P}. These QCD axions are weakly-coupled light pseudo-scalar particles generated from the spontaneously broken Peccei-Quinn $U(1)_\mathrm{PQ}$ symmetry~\cite{1978PhRvL..40..223W,1978PhRvL..40..279W,PhysRevLett.43.103,SHIFMAN1980493,DINE1981199,Zhitnitsky:1980tq}. This concept can be generalized to a class of light pseudo scalar particles which are collectively referred to as axion-like particles~(ALPs). They are motivated by a spontaneously broken global $U(1)$ symmetry beyond the Standard Model~(SM), such as those appearing in string theory~\cite{Svrcek_2006,PhysRevD.81.123530,ringwald2014axions}.
The generalized ALP field $a$ can include non-gravitational couplings arising from the following interaction Lagrangian
\begin{equation}
    \mathcal{L}_\mathrm{int}=\frac{\partial_\mu a}{f_\mathrm{lin}}\bar\psi\gamma^\mu\gamma^5 \psi+\frac{\partial_\mu a^2}{f_\mathrm{quad}^2}\bar\psi\gamma^\mu\gamma^5 \psi+\cdots\label{eq:interaction_lagrangian},
\end{equation}
where $f_\mathrm{lin}$ and $f_\mathrm{quad}$ are the effective linear and quadratic interaction scales in energy units, $\psi$ is a fermion field in SM~\cite{2013PhRvL.110b1803P,2016PhR...643....1M,Choi_2021}. This allows ALPs to couple to atomic spins through a gradient interaction.

In most direct searches for dark matter, the density of dark matter in the solar system is assumed to be relatively uniform~\cite{doi:10.1146/annurev-astro-081817-051756}. However, in addition to this conventional model of dark matter distribution, it is possible that the local dark matter density is highly nonuniform. This can occur as a result of the formation process of pseudo-scalar fields during cosmological inflation~\cite{1982PhRvL..48.1156S,VILENKIN1985263}. 
An example is the Kibble mechanism~\cite{Kibble_1976} which describes a cosmological phase transition during the cooling down of the early Universe. The phase transitions associated with symmetry breaking might induce local selections of broken symmetry and eventually result in separated domains with locally degenerate broken symmetry. Then, this can naturally lead to topological defects if the separation between domains are too far to communicate.
The type of defect mainly depends on the property of the broken symmetry and the characteristics of the phase transition but can be classified based on their dimensionality: monopoles in 0D, strings in 1D, and domain walls~(DWs) in 2D or higher dimensions. Among them, the DWs are objects formed from the discrete broken symmetry at the phase transition, and a network of such DWs may divide the universe into different sections. The size of DWs is assumed to be on the scale of $d\approx 1/m_a$ where $d$ is the thickness of the DW and $m_a$ is the mass of the ALP.

DWs may contribute to the dark matter in the universe. However, stable DWs of QCD axions would be cosmologically disastrous because they would store too much energy~\cite{1975JETP...40....1Z,PRESKILL1991207}.
Nevertheless, ALP DW could exist up to the modern epoch in the post-inflation scenario since ALP fields are not restricted by the QCD phenomology
~\cite{PhysRevD.55.5129,Hiramatsu_2013,PhysRevD.101.023514,PhysRevD.101.035029}. 
%Self-interacting ALP can also be composite object dark matter such as axion miniclusters or boson stars~\cite{PhysRevD.101.083014,PhysRevLett.117.121801,PhysRevD.97.043002,PhysRevLett.71.3051}. 
If they indeed exist, such ALP DW dark matter would have a highly nonuniform local density.

Recently, a series of experiments have been proposed to attempt the direct detection of locally dense dark matter~\cite{2013PhRvL.110b1803P,2013AnP...525..659P,PhysRevX.4.021030,Roberts2017Search-for-doma,PhysRevLett.121.061102,Garconeaax4539, McNally2020Constraining-do,Figueroa_2021}.
The Global Network of Optical Magnetometers for Exotic physics searches~(GNOME) is the first experiment to look for localized dark matter governed by transient spin-dependent interactions between the dark matter and atomic spins~\cite{2013AnP...525..659P,PhysRevD.97.043002,AFACH2018162,MASIAROIG2020100494}. Details of this experiment are described in Refs.~\cite{2013AnP...525..659P,AFACH2018162}.

The common idea of these experiments is to distribute multiple sensors (optical magnetometers in the case of GNOME) across geographically separated locations on the Earth, and connect them as an array. The passage of the Earth through a localized dark matter object may cause an interaction at each sensor with a distinctive amplitude at a certain time depending on the spatiotemporal distribution of the dark matter.

However, the behavior of the signal amplitude and timing from a dark matter crossing event is a priori unpredictable since the local density of the dark matter cannot be determined from existing theories. Therefore, the analysis of such measurements needs to have a feasible model to determine whether the signal pattern actually corresponds to a possible dark matter crossing event. 
For example, if the distance between domains $L$ is much larger than the Earth scale ($L\gg R_\oplus$), the boundaries of such DW would be approximated as a flat object with non-zero thickness~\cite{2013PhRvL.110b1803P}.
Then the DW crossing events captured by the detector network can be described by a simple parametric template with the relative velocity and orientation of a flat DW. Nevertheless, the measurements may contain a large multidimentional array of information which could cause complications when accessing from a conventional data analysis scheme.

Recent developments in machine learning~(ML) techniques have shown the ability to extract a particular feature from multi-dimensional datasets in various fields, including physics~\cite{Goodfellow-et-al-2016,RevModPhys.91.045002}. One of these ML methods, stochastic optimization (or stochastic gradient descent) with adaptive momentum, enables one to fit a parametric template of crossing events in a time window of network data without massive calculations of every possible combination of templates~\cite{QIAN1999145,JMLR:v12:duchi11a,tieleman2012lecture,zeiler2012adadelta,2014arXiv1412.6980K}.
This ML assisted fitting method can be used for discerning whether the network data can be well fit to any of the possible DW crossing events. This optimization can also be extended to search not only for DWs, but also for a variety of other locally dense dark matter objects~\cite{PhysRevD.97.043002,PhysRevLett.71.3051,PhysRevLett.117.121801,PhysRevD.101.083014,Dailey2021Quantum-sensor-}.

In this paper, a data analysis method to discern DW crossing events from the GNOME data is presented. This analysis scheme utilizes a parametric template of DW crossing events, instead of scanning a possible parameter lattice (as was done in Refs. \cite{MASIAROIG2020100494,afach2021search}). The feasible parameter range of the template and event detection threshold are optimized via a simulated dataset.

The expected signal amplitudes and timings of the DW crossing event are derived in Sec.~\ref{sec2}. The procedure of the data analysis is presented in Sec.~\ref{sec3}. The reliability and validity of the data analysis are studied based on binary classification as described in Sec.~\ref{sec4}. The conclusion and prospects for future data analysis are described in Sec.~\ref{sec5}.

%%%%%%%%%%%%

\section{Parametric template of the domain-wall crossing event\label{sec2}}
The ALP fields inside the DW between two neighboring vacua along the normal direction parametrized by $z$ can be described as~\cite{2013PhRvL.110b1803P,afach2021search}
\begin{equation}
a(z)=\frac{4f_\mathrm{SB}}{\sqrt{\hbar c}}\arctan\left(\exp\left(\frac{m_ac^2}{\hbar c}z\right)\right),
\end{equation}
where $f_\mathrm{SB}$ is the symmetry-breaking scale of the ALP.
The pseudoscalar linear coupling of the ALP field $a$ to the Standard Model axial-vector current has the interaction Hamiltonian from Eq.~(\ref{eq:interaction_lagrangian}) as
\begin{equation}
H_\mathrm{int}=\frac{\left(\hbar c\right)^{3/2}}{f_{\mathrm{int}}}\hat S\cdot\vec\nabla a,
\end{equation}
where $f_{\mathrm{int}}$ is the effective interaction scale between the gradient of the ALP field $\vec\nabla a$ and unit fermionic spins $\hat S$. The effective interaction scale can be different for electrons, neutrons, and protons depending on the particular theoretical model, but here we consider only ALP-proton coupling. The GNOME magnetometers are inside multi-layer magnetic shield, which cancel the effects due to electron spin couplings due to the an induced magnetization of the shield~\cite{PhysRevD.94.082005}. All GNOME magnetometers use atoms whose nuclei have valence protons, and thus are primarily sensitive to ALP-proton interactions~\cite{Kimball_2015}.

The ALP-DW interactions with atomic spins can be interpreted as a pseudo-magnetic field $\vec B$ acting on the atoms.
It can be written in analogy with the Zeeman Hamiltonian form as
\begin{equation}
H=-\gamma\vec S\cdot\vec B,
\end{equation}
where $\gamma$ is the gyromagnetic ratio of atomic species employed in each magnetometer. While a conventional magnetic field is screened by the multi-layer magnetic shields, the ALP-proton coupling is not screened, and can be detected by the atomic magnetometer.

The expected strength of the pseudo-magnetic field $B_s$ at the magnetometer labeled by $s$ is derived as
\begin{equation}
B_s=\frac{4}{\mu_B}\frac{f_\mathrm{SB}}{f_\mathrm{int}}m_ac^2\frac{\sigma_s}{g_{F,s}}\cos\psi_s,
\end{equation}
where $\mu_B$ is the Bohr magneton, $\sigma_s/g_{F,s}$ is the estimated ratio between the effective proton spin polarization and the Land\'e $g$-factor for the magnetometer $s$, and $\psi_s$ is the angle between the ALP-field gradient and the sensitive direction of the magnetometer~\cite{afach2021search}.
Here, $f_\mathrm{SB}$ is constrained by the cosmological parameters as
\begin{equation}
f_\mathrm{SB}=\hbar c\sqrt{\frac{L\rho_\mathrm{DW}}{8m_ac^2}}\leq\hbar c\sqrt{\frac{L\rho_\mathrm{DM}}{8m_ac^2}},
\end{equation}
where $\rho_\mathrm{DW}$ and $\rho_\mathrm{DM}$ are the energy density of the DW and dark matter, respectively. 
The expected strength of the pseudo-magnetic field can be written with cosmological parameters $L$ and $\rho_\mathrm{DW}$ as
\begin{equation}\label{eq:mag_t}
B_s=\frac{4}{\mu_B}\frac{\sigma_s}{g_{F,s}}f_\mathrm{SB}\sqrt{m_ac^2}\frac{\sqrt{m_ac^2}}{f_\mathrm{int}}\cos\psi_s=\frac{4\hbar c}{\sqrt{8}\mu_B}\frac{\sigma_s}{g_{F,s}}\sqrt{L\rho_\mathrm{DW}}\frac{\sqrt{m_ac^2}}{f_\mathrm{int}}\cos\psi_s.
\end{equation}
We assume that $L\approx7.5\times 10^{-5}~\mathrm{ly}$ (wall crosses the Earth approximately once a month with a relative velocity $\left|v_d\right|=10^{-3}c$) and $\rho_\mathrm{DW}=0.4~\mathrm{GeV}/\mathrm{cm}^3$ (local dark matter density~\cite{Read_2014}). Independently measured $L$ and $\rho_\mathrm{DW}$ can change the effective interaction scale.

The expected amplitude of the effective magnetic field $B_s$ depends not only on the ALP parameters (mass and effective interaction scale) but also on the properties of the magnetometer and geometric factors. In particular, the factor $\cos\psi_s$ that describes a relative crossing direction could suppress the magnetometer signal, even to zero in some cases. 

If a DW crossing event happens, the timings $t_{0,s}$ of the expected signal at each detector $s$ are also unpredictable in a priori because they are determined by the relative positions between the magnetometers and DW.
For simplicity, we assume that the DW is relatively flat with respect to Earth scale, and it travels at velocity $\vec v_d$, initially at $\vec x_d$ as the closest point in the wall to Earth.
Then the encounter time can be estimated as
\begin{equation}\label{eq:timing}
t_{0,s}=\frac{\left(\vec x_d-\vec x_s\right)\cdot\hat x_d}{\left|\vec v_d\right|},
\end{equation}
where $\vec x_s$ is the position of the magnetometer $s$.
The signal duration $\tau$ is characterized by the DW thickness $d=2\sqrt{2}\hbar c/m_ac^2$ and the relative speed $\left|\vec v_d\right|$
\begin{equation}\label{eq:duration}
\tau=\frac{2\sqrt2\hbar c}{m_ac^2\left|\vec v_d\right|},
\end{equation}
independent of the sensor.

\begin{figure}
\includegraphics[width=0.5\textwidth]{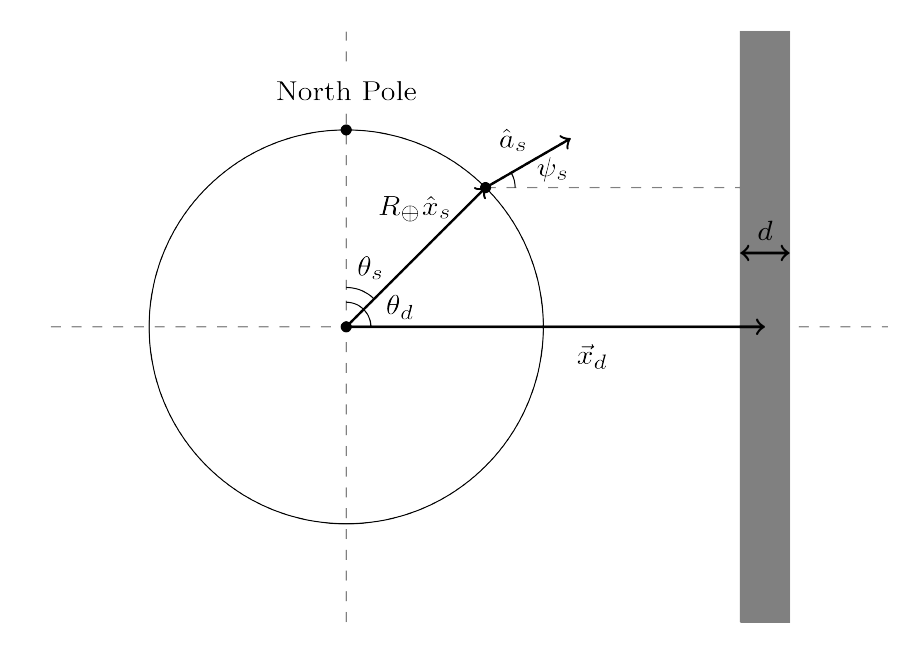}
\caption{Conceptual diagrams of the Earth (left circle) and DW (right rectangle) are represented to describe the geometrical parameters. For each magnetometer $s$, its position vector $\vec x_s=R_\oplus\hat x_s$ is described by the spherical coordinates $(R_\oplus,\theta_s,\phi_s)$ centered at the Earth center. The sensitive direction vector of the magnetometer $s$ is denoted by $\hat\alpha_s$. A position of a DW, $\vec x_d$, is described in the same manner as $(|\vec x_d(t)|, \theta_d,\phi_d)$. $d$ is the DW thickness and $\psi_s$ is the angle between $\hat\alpha_s$ and $\vec x_d$. \label{fig:geometry_notation}}
\end{figure}

Figure~\ref{fig:geometry_notation} shows the conceptual diagram of the magnetometer $s$ on the Earth's surface at $\vec x_s=R_\oplus\hat x_s$ with the sensitive direction $\hat\alpha_s$ and the DW at its position $\vec x_d$. $R_\oplus$ denotes the Earth radius. Each of the position vectors can be represented using spherical coordinates centered at the Earth's center, aligned to the North pole (polar angle) and prime meridian (azimuthal angle). 
Then the DW crossing event parameters for generating the signal pattern (amplitude $B_s$, timing $t_{0,s}$, and duration $\tau$) are determined by the six parameters listed in Table~\ref{tab:DWparams_norm}.

%%%%%%%%%%%%%%%
% SECTION III %
%%%%%%%%%%%%%%%

\section{Method\label{sec3}}
%A geographically correlated pattern, like that due to a DW crossing event, can be evaluated with respect to how well does such a pattern fit the GNOME data.
The likelihood of a geographically correlated signal pattern in the data being produced by a DW is evaluated by comparing it to the predicted pattern from Eqs.~(\ref{eq:mag_t}), (\ref{eq:timing}), and (\ref{eq:duration}). If such an event occurs and generates signals distinguishable from noise, the relevant physical parameters of the correlated pattern can also be estimated. This process can be divided into several steps. 
(1) Prepare a test-time window to be analyzed containing data from the stations operating during that specific interval of time.
Data pre-processing is applied to each point in the test-time window. 
(2) Generate the parameter space for variables used to optimize the parametric model for the data. 
(3) Perform the stochastic optimization for fitting the pattern to the data and evaluate the goodness of the fitting. Here, the parameter-estimation error will be used. 
(4) Characterize the test-time window based on the evaluated estimation error.

A test-time window of data is defined by a set of discrete data points from each sensor in the network. For a given time interval and sampling rate, the test-time window is constructed from all data points from all available sensors during the time interval. The linear baseline of each sensor is removed by subtracting a linear fit to the data in advance. There are no additional filters or time binning in the pre-processing step.

Then the stochastic optimization iteratively updates the template parameters to fit a signal pattern to the test-time window. In order to build the stochastic optimization process to search for DW crossing events, the proper template parameters of the event should be determined. The parameters are updated based on the gradient with respect to them in the parameter space, toward the minimum value of the estimation error. 
Each parameter should be normalized to prevent a directional bias during every updating iteration. 
This normalization requires a definite boundary for each parameter.

%%%%%%%%%%%%%%%%%%%
%%% Section III. A
%%%%%%%%%%%%%%%%%%%

\subsection{Boundary, distribution, and normalization of the DW crossing event parameters}
For ALP DW, the mass $m_ac^2$ and the effective interaction scale $f_\mathrm{int}$ are free parameters. So they are treated as unknown parameters with logarithmic-uniform distribution. The values of $m_ac^2$ and $f_\mathrm{int}$ will be estimated as a point in the distribution during the analysis. Their boundaries are set by referring to the prior GNOME analysis range as shown in Table~\ref{tab:DWparams_norm}~\cite{afach2021search}.
%For ALP DW, the mass $m_ac^2$ and the effective interaction scale $f_\mathrm{int}$ are free parameters. So they are treated as unknown parameters with logarithmic-uniform distribution. Their boundaries are set by referring to the prior GNOME analysis range as shown in Table~\ref{tab:DWparams_norm}~\cite{afach2021search}.

The direction parameters, polar and azimuthal angles of the DW, have a feasible range with a linear-uniform distributions, where the polar angle $\theta_d$ satisfies $\theta_d\in\left[0,\pi\right]$ and the azimuthal angle $\phi_d$ satisfies $\phi_d\in\left[0,2\pi\right)$ with a periodic boundary condition.

The speed should be considered a random variable from the Maxwell-Boltzmann distribution if DWs are virialized in our galaxy. In addition, the escape velocity at a solar system galactic radius from the Milky Way's gravity is given by $v_e=550.9~\mathrm{km}/\mathrm{s}$, so we assume that faster DWs cannot exist in the galaxy~\cite{Kafle_2014}. 
A slow DW could leave a signal pattern on the data, but long-term linear drifts in the magnetometers make such a DW to be difficult to characterize~\cite{AFACH2018162}.
Since GNOME data files are stored for every minute (regardless of the data acquisition rate), the minimum speed of the DW that allows the DW to pass the Earth within at most $n$-concatenated data files is
\begin{equation}
v_\mathrm{min}=\frac{2R_\oplus}{n~\mathrm{minutes}}\approx\frac{213}{n}~\mathrm{km}/\mathrm{s}.
\end{equation}
In order to minimize the effect of drifts in the magnetometer data, $n$ is chosen to be 2. Then the boundary of the relative speed of the DW is $\left|\vec v_d\right|\in\left[100~\mathrm{km}/\mathrm{s},550~\mathrm{km}/\mathrm{s}\right]$. 
A 7~$\mathrm{km}/\mathrm{s}$ margin is given for the lower bound for the convenience of calculation. 
The speed follows the Maxwell-Boltzmann distribution with the scale parameter of $220/\sqrt2~\mathrm{km}/\mathrm{s}$ and edges on both sides are cut to exclude slow and fast DWs~\cite{PhysRevD.42.3572}.

The initial relative distance is also restricted by the same data length, $n=2$. 
To contain all signal peaks (from the closest and farthest surfaces of the Earth) in a test-time window, the boundary is needed to be
\begin{equation}
\left|\vec x_d\right|\in\left[R_\oplus,\min\left(\left|\vec v_d\right|\right)\cdot n~\mathrm{minutes}\right]
\end{equation}
The initial relative distance is uniformly distributed in the range.

In the worst case (the slowest and farthest DW from the Earth), $\left|\vec v_d\right|=107~\mathrm{km}/\mathrm{s}$ and $\left|\vec x_d\right|=12\times10^3~\mathrm{km}$, 
the DW could not pass through the Earth completely within a given test-time window of $n=2$. 
Instead, it will be contained in the one-minute overlapping neighboring test-time window with a new set of parameters $\left|\vec v_d\right|=107~\mathrm{km}/\mathrm{s}$ and $\left|\vec x_d\right|=6.4\times10^3~\mathrm{km}$.
This DW or other combinations of slow and far DWs can be searched with $n/2$-minutes overlapping adjacent test-time windows by shifting $\left|\vec x_d\right|$.
% the DW could not pass through the Earth within a given test time window of $n=2$. Instead, it will be contained in the one-minute overlapping neighboring test time window. To search for such DWs, the adjacent test time windows with overlapping $n/2$ minutes can be swept. 
For a single test-time window search, the range of the DW crossing event for the relative speed and distance $\left|\vec v_d\right|\in\left[100~\mathrm{km}/\mathrm{s},550~\mathrm{km}/\mathrm{s}\right]$ and $\left|\vec x_d\right|\in\left[6.4\times10^3~\mathrm{km},12\times10^3~\mathrm{km}\right]$ is sufficient.

The range and distribution of parameters describing the DW crossing events are represented in Table~~\ref{tab:DWparams_norm}. In actual analysis, they are calculated in the normalized unit space through normalization maps. The normalization enables the parameters with different scales and units to have a uniform gradient scale during the stochastic optimization.
%Estimating the DW parameters requires a normalization of the parameter boundary to apply stochastic optimization, which evaluates a gradient of a function in the parameter space. Such a gradient is scaled by the parameter range, so the normalized space guarantees a uniform bias for every parameter. Since two ALP parameters $m_ac^2$ and $f_\mathrm{int}$ need to be logarithmically flat, they are normalized by their power as represented in Table~\ref{tab:DWparams_norm} with other parameters.

\begin{table*}
\caption{
Six parameters describing DW crossing event and their estimation range. The parameter boundaries are normalized to the unit interval, and the azimuthal angle indicating the DW has a periodic boundary condition.
%Parameters of the DW crossing event and estimation range of each parameter. ALP field mass and effective interaction scale are normalized to a logarithmic scale.
\label{tab:DWparams_norm}}
\begin{ruledtabular}
\begin{tabular}{lcll}
Parameters & Symbols & Estimation ranges & Normalization maps $f(x):x\mapsto f(x)$\\
\colrule
mass & $m_ac^2$ & $\left[10^{-15}~\mathrm{eV},10^{-11}~\mathrm{eV}\right]$ & $(\log_{10}(x/\mathrm{eV})+15)/4$\\
interaction scale & $f_\mathrm{int}$ & $\left[10^{4}~\mathrm{GeV},10^{8}~\mathrm{GeV}\right]$ & $(\log_{10}(x/\mathrm{GeV})-4)/4$\\
polar angle  & $\theta_d$ & $\left[0,\pi\right]$ & $x/\pi$\\
azimuthal angle & $\phi_d$ & $\left[0,2\pi\right)$ & $x/2\pi$\\
relative speed & $\left|\vec v_d\right|$ & $\left[100~\mathrm{km/s},550~\mathrm{km/s}\right]$ & $(x-100~\mathrm{km/s})/450~\mathrm{km/s}$\\
relative position & $\left|\vec p_d\right|$ & $\left[6.4\times10^3~\mathrm{km},12\times10^3~\mathrm{km}\right]$ & $(x-6.4\times10^3~\mathrm{km})/5.6\times10^3~\mathrm{km}$\\
\end{tabular}
\end{ruledtabular}
\end{table*}

\subsection{Stochastic optimization}
Stochastic optimization is a stochastic version of the gradient descent optimization, which finds the minimal value of a given function based on Newton's method. A function to be optimized, the cost function, is traced by updating based on the gradient with respect to optimization parameters. In our case, the normalized optimization parameters are DW crossing event parameters listed in Table~\ref{tab:DWparams_norm}. However, since there are inevitably noise fluctuations in the data, and infinitely many forms of the signal pattern for a given DW crossing event, the shape of the cost function should reflect such fluctuations and non-linearity.

An ansatz of the cost function $\mathcal E=\frac1N\sum_s\mathcal E_s$ to estimate the DW crossing event for each magnetometer $s$ is defined as
\begin{equation}\label{eq:cost_function}
\mathcal E_s=\frac{1}{\mathcal T\sigma_s^2}\int_0^{\mathcal T}\left(\int_0^t\left(S_s(t')-\tilde S_s(t')\right)dt'\right)^2dt,
\end{equation}
where $N$ is the number of magnetometers, $\mathcal T$ is the time interval of the test-time window, $\sigma_s$ is the standard deviation of the time-series data at the magnetometer for a given time window, $S_s(t')$ is the baseline-removed time-series data of the magnetometer $s$ at time-series point $t'$, and $\tilde S_s(t')$ is the expected signal pattern of the magnetometer $s$ at time-series point $t'$. The time variables $t$ and $t'$ in $[0,\mathcal T]$ represent the common time interval for the sensors, digitized during the data acquisition.

The cumulative integration over the time-series point $t'$ can reduce the local fluctuation of the Gaussian noise. $\mathcal T$ in the denominator normalizes the length of the interval, while the variance of the magnetometer $\sigma_s^2$ normalizes the Gaussian noise. Practically, the magnetometer data does not precisely follow a Gaussian distribution, but this pseudo-normalization factor $1/\sigma_s^2$ weights the contribution of each magnetometer to the cost functions by their intrinsic noise. 
Therefore, it makes the cost function have less dependence on the sensor characteristics, or even the number of sensors in general.

\begin{figure}
\includegraphics[width=0.5\textwidth]{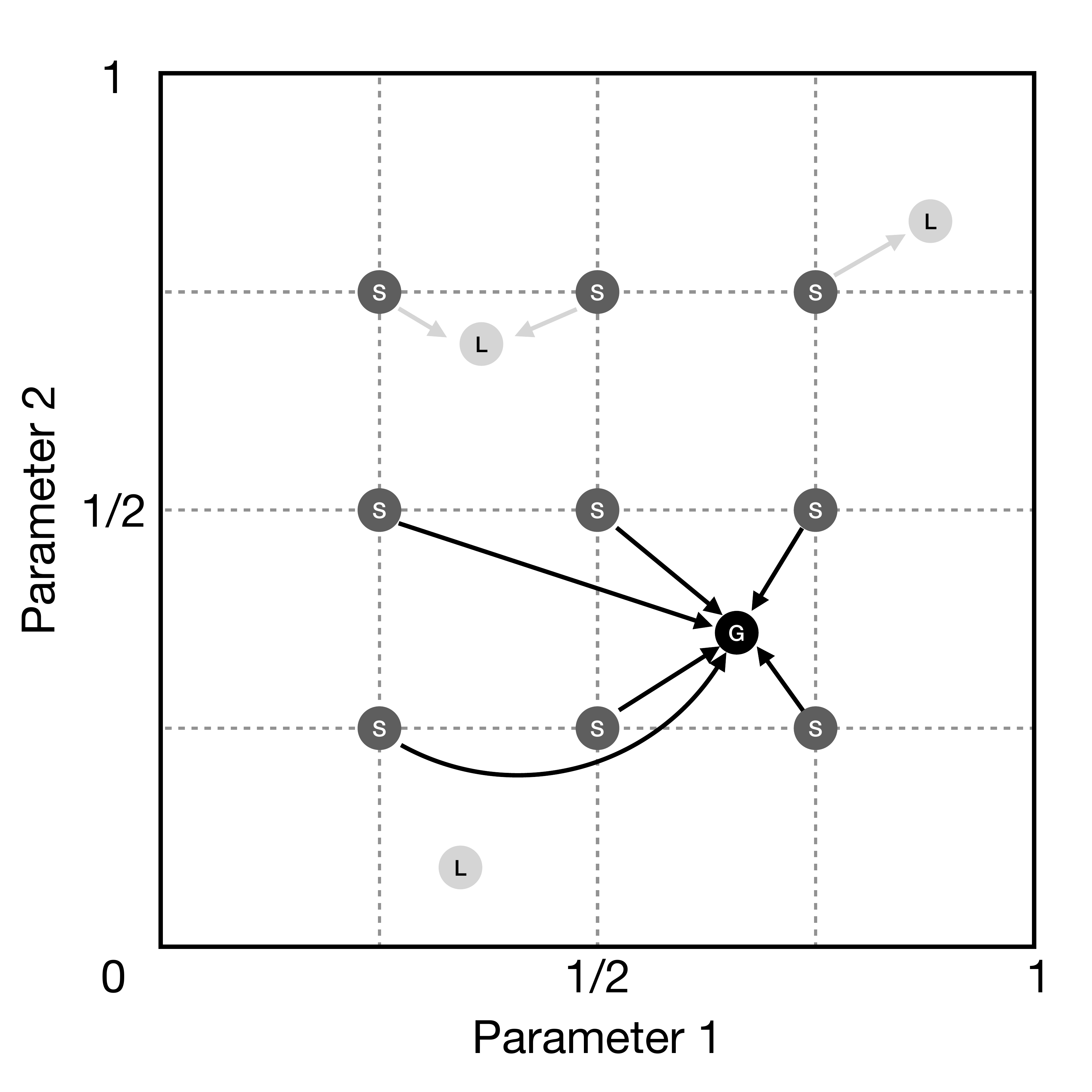}
\caption{A conceptual diagram of the grid estimation with various initial points in the normalized parameter space. In two normalized parameters estimation, there is one global minimum (G) that is the ground truth, but local minima (L) obstruct the estimation. The grid estimation uses multiple initial points (S) in the parameter space to find the most probable estimation solution.\label{fig:grid_estimation}}
\end{figure}

The remaining dependence on detector characteristics can be avoided by clustering distinct optimization processes from the multiple initial points, which generate $\tilde S_s$ on the estimating parameter space grid. Multiple initial points are defined as intersections of $D$ grid lines on the parameter space. The optimization is conducted for each $D^m$ initial points for $m$ parameters. Figure~\ref{fig:grid_estimation} shows a conceptual diagram of the grid estimation in the normalized parameter space with $D=3$. Without loss of generality, the parameter space is represented as a two-dimensional space with two normalized parameters as an example for demonstration. Then it has $3^2=9$ initial grid points.

The optimization process is conducted for each initial point (S), and each of them move towards the local minima (L) or the global minimum (G) depending on the situation. They may not converge in a finite number of iterations to any of minima. Instead, the estimated parameter points (including minima) at the last iteration step cluster in the parameter space with a distance $d_c$. Then each of them forms a hypersphere with the radius $d_c$ representing the cluster. The cost function and parameter values at clusters are averaged and compared to evaluate the minimal cost cluster. The minimal cost cluster is a representative estimated parameter point of the grid estimation.

Two parameters determine the overall optimization performance of the grid estimation. The number of grid lines $D$ balances computing time against precision. Also, a clustering distance $d_c$ among optimized results is related to accuracy.

The ADAM (ADAptive Momentum) optimization is employed for the stochastic optimization process, which can cover a larger range of the evaluated gradients by using machine learning~\cite{2014arXiv1412.6980K}. The expected signal amplitudes can vary from zero to more than a picoTesla amplitude depending on the parameters and geometric properties of the DW crossing event. The ADAM optimization can update the estimating parameters during the optimization, to fit the pattern to wide ranges of signal amplitudes with a universal process.

\subsection{Performance evaluation}
\begin{table*}
\begin{ruledtabular}
\caption{Geographic and geometric information of the GNOME station magnetometers. The magnetometer position is based on the global positioning system (GPS), where the West and South directions have a negative sign. The sensitive direction is based on the horizontal coordinate system.\label{tab:gnomelist}}
\begin{tabular}{lrrrrr}
\multirow{2}{*}{GNOME stations} & \multicolumn{2}{c}{Magnetometer position} & \multicolumn{2}{c}{Sensitive direction} & \multirow{2}{*}{$\sigma/g_F$} \\
\cline{2-3} \cline{4-5}
& Longitude [deg] & Latitude [deg] & Altitude [deg] & Azimuth [deg] &\\
\colrule
 Berkeley 1 & $-122.3$ & 37.9 & 0 & 28 & -0.39 \\
 Berkeley 2 & $-122.3$ & 37.9 & 90 & 0 & -0.39\\
 Daejeon & 127.4 & 36.4 & 90 & 0 & -0.39\\
 Hayward & $-122.1$ & 37.7 & $90$ & 0 & 0.70\\
 Krakow & 19.9 & 50.0 & 0 & 45 & 0.50\\
 Lewisburg & $-76.9$ & 41.0 & 90 & 0 & 0.70\\
 Los Angeles & $-118.4$ & 34.1 & 0 & 270 & 0.50\\
 Mainz & 8.2 & 50.0 & $-90$ & 0 & 0.50\\
 Moxa & 11.6 & 50.6 & 0 & 270 & -0.39\\
 Oberlin & $-81.8$ & 41.3 & 0 & 300 & -0.49\\
\end{tabular}
\end{ruledtabular}
\end{table*}

The stochastic optimization evaluates an estimation error given by the cost function value $\mathcal E$. If the estimation error is small enough for a given test-time window, a physical model (the DW crossing event in this case), and an appropriate optimization process, then this test-time window has a possibility to contain the physical event. 
However, noisy test-time widows can sometimes produce a small estimation error. On the other hand, a high value of $\mathcal E$ means it does not contain the physical event or the optimization process cannot find an appropriate solution. The decision about the presence of an event is characterized by a binary classification. We can define four different cases in terms of whether our data contains an event and whether the algorithm identifies it as true positive (TP), false negative (FN), false positive (FP), and true negative (TN). %Which means, whether the decision of positive (event) or negative (not event) is true or false. 

The binary classification of the DW crossing event has been tested based on simulations with GNOME data. The active magnetometers of the GNOME from January 30th to April 30th, 2020 are listed in Table~\ref{tab:gnomelist}. Since each of the magnetometers in the GNOME do not always provide proper data due to local glitches, the data may not be continuous in time. Therefore, the available dataset from magnetometers for a given time interval may have different combinations of magnetometers. For each simulation, a random combination of magnetometers is chosen to cover the most general case. However, the number of magnetometers must be larger than 3 in order to at least estimate the direction of the DW. To prevent any real dark matter event appearing during the simulation, a two-minute time window was generated using the mixed date and time of each magnetometer station.

In total 400 simulations were conducted for the binary classification. 200 simulations have a randomly injected DW crossing event from the boundary and distribution of the estimating parameters, while the remaining 200 simulations have artificially injected random noise Lorentzian peaks (0 to 4 peaks at random timings, uniformly distributed between $-50$ and $50$~pT amplitude, uniformly distributed between $0.12$ and $12$ seconds long full-width at half-maximum) to test the robustness of the algorithm against other patterns corresponding no DW crossing event. Both signal and noise peak shapes are assumed to be Lorentzian. The number of grid lines $D=2$ and the clustering distance $d_c=0.02$ were set during analysis. For each optimization at a single grid point, the DW parameters are estimated by 500 iterations steps of the ADAM optimization.

In order to characterize the binary classification, each simulation is identified as positive or negative depending on the decision criterion, which in this analysis is the estimation error $\mathcal E$. Therefore, the true positive rate (TPR) and false positive rate (FPR) are derived with respect to the estimation error. The decision criterion threshold $\mathcal E_\mathrm{th}$ is determined at 95\% level of the TPR. The confidence interval of each rate $p$ (TPR or FPR) in $n$ observations is described by the continuity-corrected Wilson score interval with a critical value $z$~\cite{10.2307/20022840,https://doi.org/10.2307/2983604,Newcombe1998Two-sided-confi,doi:10.1080/09296174.2013.799918}

\begin{align}
\nonumber p&=\frac{2n\hat p+z^2}{2\left(n+z^2\right)}\\
&~\pm\frac{1}{2\left(n+z^2\right)}\left(1+z\sqrt{4n\hat p(1-\hat p)\mp2\left(1-2\hat p\right)-\frac1n+z^2}\right).
\end{align}
For the 95\% of confidence level, $z=1.96$. The continuity-corrected Wilson score interval at the 95\% confidence level would be applied to estimate the confidence intervals of TPR and FPR.

\subsection{Parameter space optimization}
The set of virtual observations, 400 simulations, is classified within the parameter space bounded by the estimation range listed in Table~\ref{tab:DWparams_norm}. In the presence of an event for a given test-time window, the optimization estimates the corresponding DW parameters simultaneously. The estimated parameters need to be accurately converged, but sometimes they do not converge to stable values within a reasonable iteration time, due to computing limitations. This can be handled by applying a finer grid estimation and larger iterations of the fitting. 

More efficiently, the ALP field parameter space generated from the mass and effective interaction scale can be optimized without any increase in computing cost by using a two-step process. In the virtual observations with DW crossing events, which are characterized by a certain set of DW parameters, the distance between the estimated parameters and the desired parameters in the ALP field parameter space can be different for distinct regions in the space. Since the optimization is conducted in the normalized space, the distance can be defined as a norm. Then the distance indicates the error level of the estimated parameters from the desired parameters.

Let a well-estimated candidate be when the optimization gives a distance of less than 0.02 (in the normalized space). This distance and corresponding candidate are meaningful only if the analysis method detects an event, i.e., TP event cases. The accuracy is defined as a population ratio between the well-estimated candidates and the total TP events. The ALP field parameter space can be optimized to maximize the sum of the accuracy and the area of optimized space. The subspace of the ALP field parameter space is swept by the mass and effective interaction scale. Then the parameter space sensitive to this analysis method will be derived.

\section{Result\label{sec4}}
\begin{figure}
\includegraphics[width=0.45\textwidth]{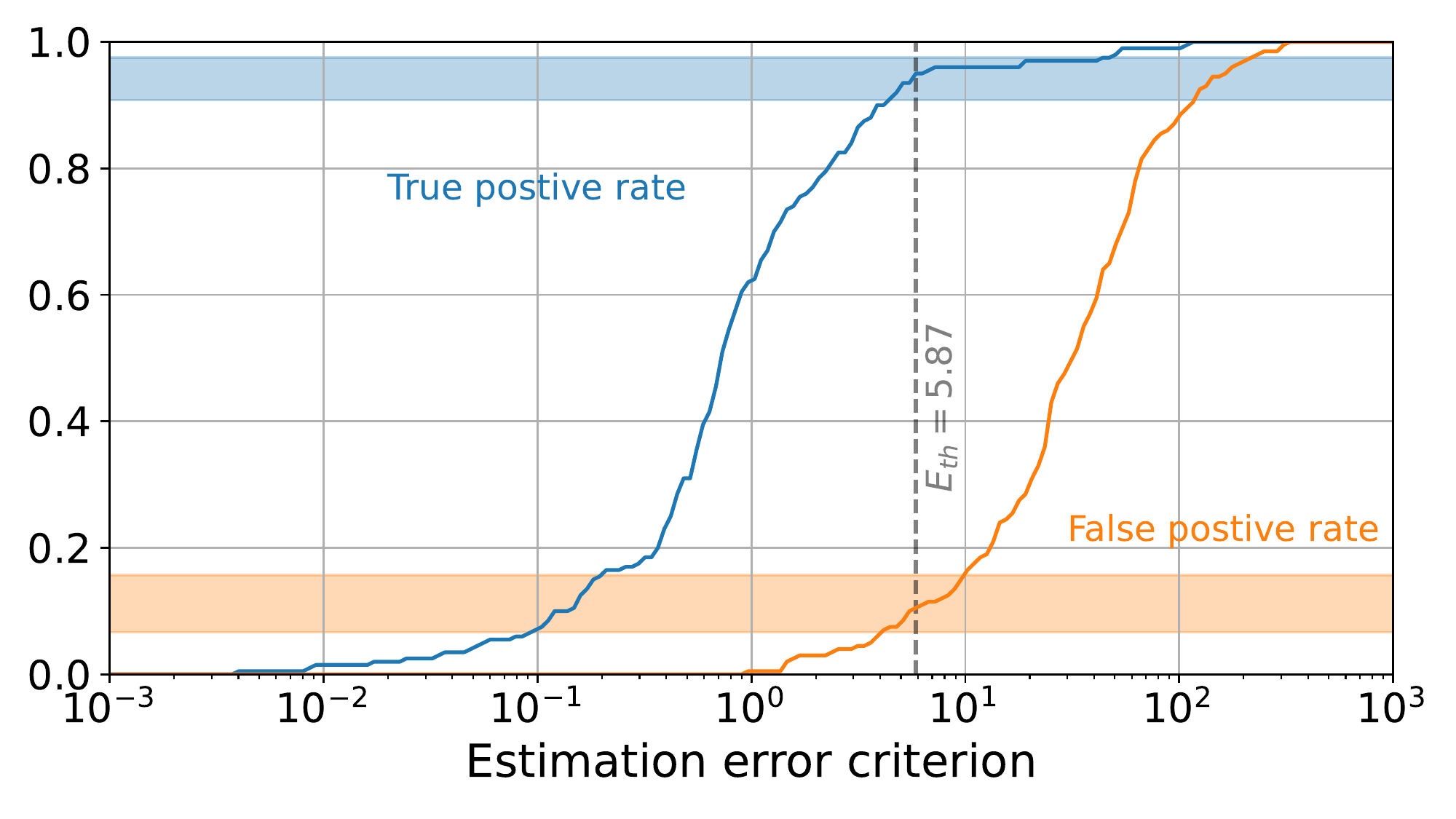}
\includegraphics[width=0.45\textwidth]{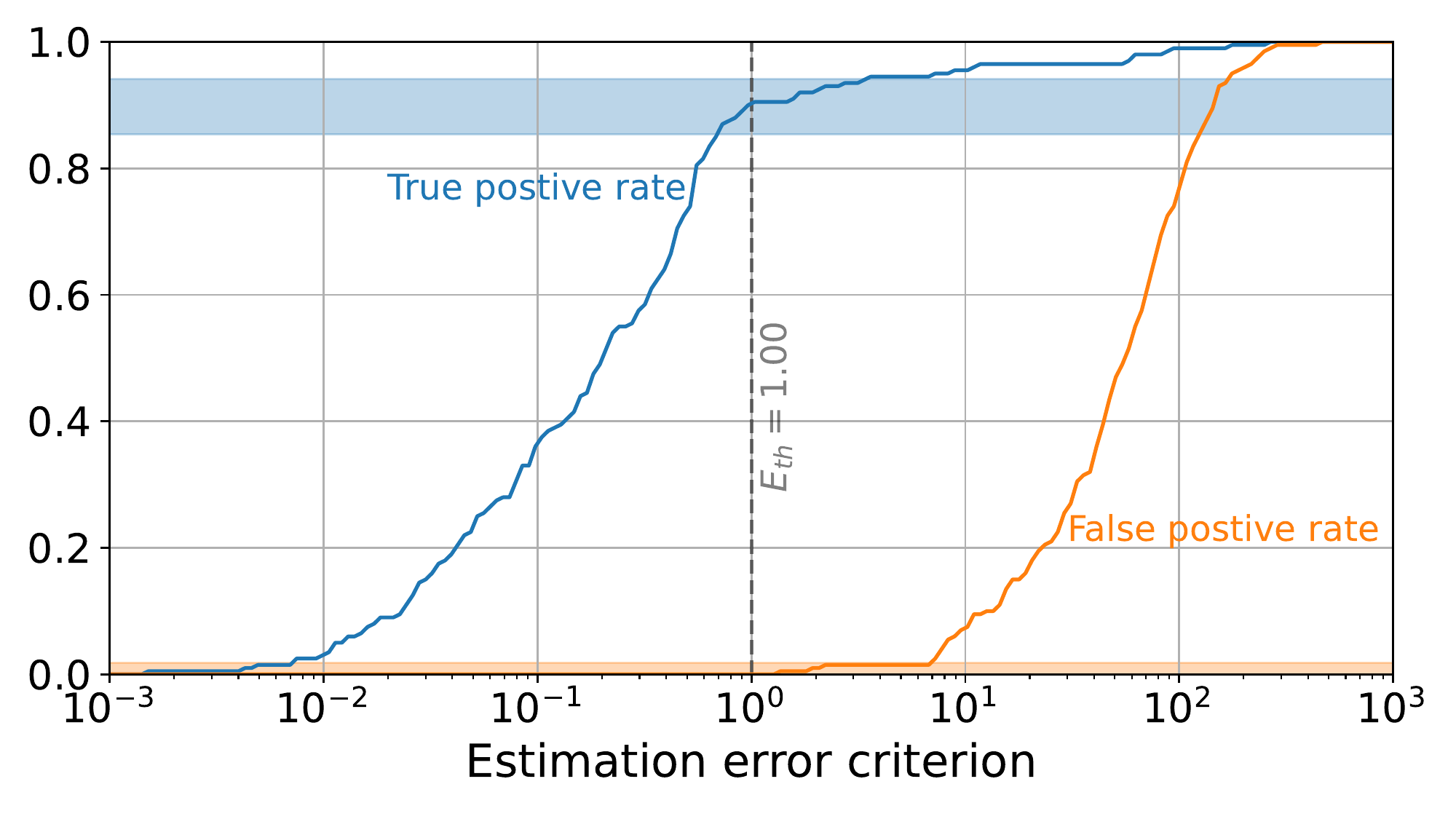}
\caption{Binary classifications of the analysis method with simulated DW crossing events before (left) and after (right) the parameter space optimization. TPR (blue) and FPR (orange) are represented with respect to the estimation error. Before the parameter space optimization, the TPR and FPR are 95\% and 10\% at the decision criterion threshold $E_\mathrm{th}=5.87$. 
The TPR and FPR are changed to 91\% and 0.0\% at $E_\mathrm{th}=1.00$ after the parameter space optimization.\label{fig:tpr_fpr}}
%The TPR and FPR are changed to 90.5\% and 0.0\% at $E_\mathrm{th}=1.00$ after the parameter space optimization.\label{fig:tpr_fpr}}
\end{figure}

Figure~\ref{fig:tpr_fpr} represents the TPR (blue) and FPR (orange) of the DW crossing event classification from the simulations before (left) and after (right) the parameter space optimization. The corresponding areas under the receiver operating characteristic (AUROC) curves were measured to be 0.963 and 0.974~\cite{PMID:3753562}. The performance of the classification was enhanced after the parameter space optimization. Before the parameter space optimization, the decision criterion threshold is derived to the value $E_\mathrm{th}=5.87$, which corresponds to the 95\% of TPR from simulations. The FPR at the threshold was observed to be 10\%. The corresponding confidence intervals with this classification algorithm were a TPR in between 90\% to 97\% and an FPR in between 6\% to 15\%.

\begin{figure}
\includegraphics[width=0.75\textwidth]{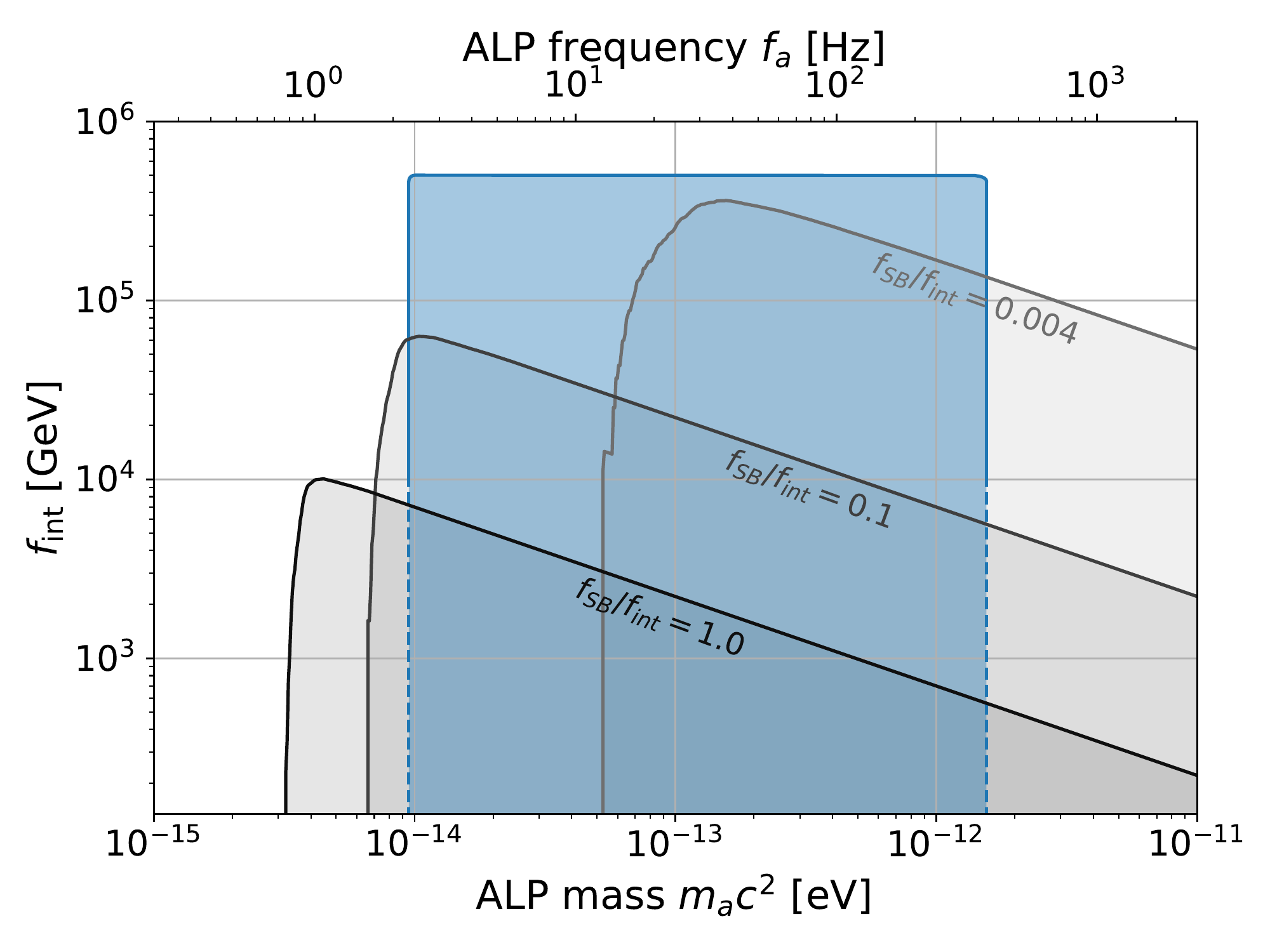}
\caption{The projected limit of the optimized parameter space identified and characterized by this method (blue). Previously analyzed regions with GNOME data and analysis method are presented for comparison (gray-scale)~\cite{afach2021search}. 
The acceptance error is 2\% for the estimated parameter within $m_ac^2$ from $1.00\times10^{-14}~\mathrm{eV}$ to $1.34\times10^{-12}~\mathrm{eV}$ and $f_\mathrm{int}$ from $10^{4}~\mathrm{GeV}$ to $4.65\times10^{5}~\mathrm{GeV}$.\label{fig:param_space}}
%The acceptance error is 2\% for the estimated parameter within $m_ac^2$ from $1.41\times10^{-14}~\mathrm{eV}$ to $1.91\times10^{-12}~\mathrm{eV}$ and $f_\mathrm{int}$ from $10^{4}~\mathrm{GeV}$ to $3.89\times10^{5}~\mathrm{GeV}$.\label{fig:param_space}}
\end{figure}

Based on the classification result, the parameter space searched as described in the Table~\ref{tab:DWparams_norm} was optimized, as shown in Figure~\ref{fig:param_space}. The 400 virtual observations were simulated within the optimized parameter subspace. 
The result shows a TPR and FPR as 91\% and 0.0\%, respectively. The TPR is in $\left[85\%,94\%\right]$ and the FPR is in $\left[0.0\%, 1.8\%\right]$ at the 95\% confidence level, when the threshold is set to 1.00 as shown in the right of Figure~\ref{fig:tpr_fpr}. 
The optimized parameter space is a projected limit of the described algorithm for the GNOME setup listed in the Table~\ref{tab:gnomelist} to search for DW crossing events, corresponding to 
$m_ac^2$ from $1.00\times10^{-14}~\mathrm{eV}$ to $1.34\times10^{-12}~\mathrm{eV}$ and $f_\mathrm{int}$ from $10^{4}~\mathrm{GeV}$ to $4.65\times10^{5}~\mathrm{GeV}$ with a 2\% acceptance error (a distance of 0.02 in the normalized unit space).
For $f_\mathrm{int}\leq10^4~\mathrm{GeV}$, the signal pattern would show the same pattern, but enlarged amplitude. They could be covered by extending the boundary of the effective interaction scale.
%$m_ac^2$ from $10^{-13.85}~\mathrm{eV}$ ($1.41\times10^{-14}~\mathrm{eV}$) to $10^{-11.72}~\mathrm{eV}$ ($1.91\times10^{-12}~\mathrm{eV}$) and $f_\mathrm{int}$ from $10^{4.00}~\mathrm{GeV}$ to $10^{5.59}~\mathrm{GeV}$ ($3.89\times10^{5}~\mathrm{GeV}$). 
This result improves the parameter space analyzed with GNOME data as described in Ref.~\cite{afach2021search}, but it is worth noting that the network status is different.
Also, this method is independent of $f_\mathrm{SB}/f_\mathrm{int}$, instead it has to scan the continuous data for estimating the domain size $L$.
%Also, this method is independent of $f_\mathrm{SB}/f_\mathrm{int}$.
The projected parameter space can be further improved as more virtual observations are simulated with intensive computations.

\section{Conclusions\label{sec5}}
A new data analysis method based on machine-learning assisted stochastic optimization has been presented to search for direct detection of localized dark matter using GNOME data. The identification and characterization of the ultra-light ALP DW events were tested in simulated virtual observations. The identification evaluated TPR with at least 85\% and FPR at most 2\% with a 95\% confidence level, and characterization showed 88\% accuracy. This accuracy can be improved by rescanning and complementary analyses.

This new method allows us to investigate signal patterns for localized dark matter using a geographically distributed network of sensors. This is not limited to DW signals only, but can also be extended to more general signals. Furthermore, it is also possible to employ a network of heterogeneous detectors if the pattern can be predicted theoretically for each detector~\cite{Dailey2021Quantum-sensor-}.

\subsection*{Data availability}
The datasets used in the current study are available from the corresponding authors on reasonable request. See also collaboration website \texttt{https://budker.uni-mainz.de/gnome/} where all the data available can be displayed.

\begin{acknowledgments}
The authors thank to Vincent Dumont and Chris Pankow for early contribution to the data analysis, and to all the members of GNOME collaboration for helpful insights and discussions. 

This work was supported by the Institute for Basic Science under grant No. IBS-R017-D1-2021-a00. 
The work of Derek F. Jackson Kimball was supported by the U.S. National Science Foundation under grant No. PHY-1707875 and PHY-2110388.
The work of Dmitry Budker was supported by the European Research Council under the European Union's Horizon 2020 Research and Innovative Program under Grant agreement No. 695405, the Cluster of Excellence ``Precision Physics, Fundamental Interactions, and Structure of Matter'' (PRISMA+ EXC 2118), DFG Reinhart Koselleck (Project ID 390831469), Simons Foundation, and Heising-Simons Foundation.
\end{acknowledgments}

\newpage

\appendix
\section{Pseudocode of the data analysis algorithm}
The pseudocode for the algorithm described in the main text is written in Algorithm \ref{alg:epe}.

\begin{algorithm}
\SetAlgoLined
\KwData{Mutiple time series $S_s(t)$ for $s\in S$}
\KwResult{Estimated error $\mathcal E=1/|S|\sum_s\mathcal E_s$}
%  initialization\;
\For{each grid point $g$}{
Optimize the domain-wall parameters $\vec d^g$\;
Evaluate estimation error $\mathcal E^g$\;
}
\For{each grid point $g$}{
 \For{each grid point $g'$}{
  \If{$\sqrt{\left|\vec d^g-\vec d^{g'}\right|^2}<d_c$}{
   Clustering $g'$ into cluster $g$\;
  }
 }
 \eIf{No clustering for any $g'$}{
 Discard the grid point $g$\;
 }{
 Evaluate representative estimation error $\mathcal E^g$ and corresponding $\vec d^g$\;
 }
}
Find $\min_g\left(\mathcal E^g\right)$ and corresponding $\vec d^g$\;
\caption{Event parameters estimation\label{alg:epe}}
\end{algorithm}

\newpage

\bibliographystyle{apsrev}
%\bibliography{}

\end{document}